\documentstyle[twocolumn,epsf]{jpsj}

\def\runtitle{
Enhancement of de Haas-van Alphen Oscillation due to Spin in 
the Magnetic Breakdown System 
}
\def\runauthor{Keita {\sc Kishigi}, 
Yasumasa {\sc Hasegawa} and 
Mitake {\sc Miyazaki}
}

\title{
Enhancement of de Haas-van Alphen Oscillation due to Spin in 
the Magnetic Breakdown System 
}

\author
{
Keita {\sc Kishigi},\footnote{E-mail: kishigi@sci.himeji-tech.ac.jp}
Yasumasa {\sc Hasegawa} and 
Mitake {\sc Miyazaki}
}

\inst
{
Faculty of Science, Himeji Institute of Technology, Akou-gun, Hyougo
678-1297, Japan \\
}

\recdate
{
\today
}

\abst
{
The effects of the Zeeman term on the de Haas-van Alphen oscillation 
is studied in the magnetic breakdown system. 
We find that the amplitude of the oscillation with the frequencies of 
$f_{\beta}$+$f_{\alpha}$ and $f_{\beta}$+2$f_{\alpha}$ are enhanced 
by the Zeeman term, while they are expected to be reduced in the 
semiclassical theory. 
A possible interpretation of the experiments in organic conductors 
is discussed. 
}

\kword
{
magnetic breakdown, 
quantum interference oscillation, spin, 
quantum magnetic oscillation, quasi-two-dimensional organic conductors
}

\begin{document}
\sloppy
\maketitle
\section{Introduction}
The magnetic breakdown\cite{Shoenberg84} is observed 
in the Shubnikov-de Haas oscillation (for the transport quantity) 
and 
the de Haas-van Alphen oscillation (for the thermodynamics quantity). 
These oscillations are so-called the quantum magnetic oscillation, 
which is attributed to the Landau quantization in the closed orbit. 
In quasi-two-dimensional organic conductors such as 
$\kappa$-(BEDT-TTF)$_2$Cu(NCS)$_2$ and 
$\alpha$-(BEDT-TTF)$_2$MHg(SCN)$_4$ (M=K, Rb, Tl and NH$_4$), 
many experiments for the magnetic breakdown are done, 
since those materials have the suitable Fermi surface (Fig. 1) for 
the study of the magnetic breakdown.\cite{review,wos} 
In this case, 
the electron orbital motion due to 
magnetic fields ($H$) applied perpendicular to $k_x$-$k_y$ plane 
is confined to the small closed orbit named as the
$\alpha$ orbit and a pair of open orbits at low fields 
as shown in Fig. 1. 
Then, this closed orbital motion leads to the Landau quantization, so that 
the magnetization and magnetoresistance oscillate by 
the frequency ($f_{\alpha}$) 
of the 
area of the $\alpha$ orbit as a function of $1/H$. 
The high field enables electrons to tunnel the Brillouin zone
gap and to move along the larger
closed orbit named as the $\beta$ orbit. Thus, 
the oscillation frequency
($f_{\beta}$) of the
area of the $\beta$ orbit gradually appears as $H$ increases. 
This is the magnetic breakdown, 
which has been explained 
in a semiclassical theory by 
Falicov and Stachoviak\cite{Falicov66} based on Pippard's 
network model.\cite{Pippard62}

In Falicov and Stachoviak theory, 
the oscillations with various frequencies ($f_{\alpha}$, 
$f_{\beta}$, $f_{2\beta}=2f_{\beta}$, 
$f_{\beta+\alpha}=f_{\beta}+f_{\alpha}$, etc.) are allowed, 
since the closed orbital motion is possible. 
However, the orbit named as the $\beta$-$\alpha$ orbit in Fig. 1 is prohibited, 
because the $\beta$-$\alpha$ orbit is not allowed to execute the 
cyclic motion in the semiclassical picture. 
The oscillation corresponding to the 
area of prohibited orbits such as the $\beta$-$\alpha$ orbit is 
called the quantum interference oscillation. 
Shiba and Fukuyama\cite{Shiba} 
have shown 
the existence of the quantum interference oscillation 
in the transport 
in the case of the magnetic breakdown. It is now well-known 
as the Stark quantum  
interference oscillation\cite{Stark77}. 
On the other hand, it was believed that 
the quantum interference oscillation in the thermodynamics 
quantity dose not exist.

\begin{figure}
\leavevmode
\epsfxsize=8cm
\epsfbox{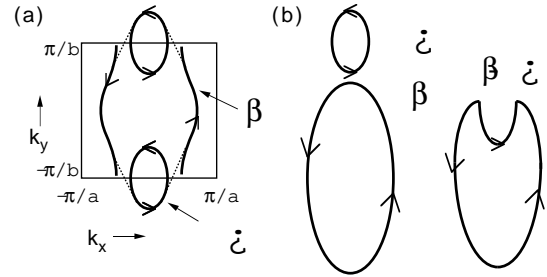}
\caption{(a) The solid lines are the Fermi surface in our calculations. 
The arrow indicates the direction of the 
electron orbital motion. 
For example, the effective masses 
for the $\alpha$ and $\beta$ orbits 
in (BEDT-TTF)$_2$$X$ are 
$m_{\alpha}/m_0\simeq 1.4$ and $m_{\beta}/m_0\simeq 3.8$, 
respectively.\cite{review,wos} 
(b) Their figures indicate the various orbits.}
\label{fig:1}
\end{figure}

However, 
by using the simple tight-binding model
,~\cite{kishigi,kishigi2,kishigi3} 
one of the authors first found 
the quantum interference oscillation, that is, 
the $\beta$-$\alpha$ oscillation in the
magnetization $M(H,n)$, whose frequency
($f_{\beta -\alpha}=f_{\beta}-f_{\alpha}$) corresponds to the area of
the $\beta$-$\alpha$ orbit in Fig. ~\ref{fig:1} 
by 
calculating the free energy, $E(H,n)$, full-quantum-mechanically at the 
fixed total electron number, $n$.

Recently, 
the existence of the $\beta$-$\alpha$ oscillation in the
magnetization was 
experimentally observed 
in $\kappa$-(BEDT-TTF)$_2$Cu(NCS)$_2$\cite{Meyer,uji} and 
$\alpha$-(BEDT-TTF)$_2$KHg(SCN)$_4$.\cite{honold} 

From other aspects of theory,
\cite{nakano,alex,harrison,sandu,so} 
the existence of the 
$\beta$-$\alpha$ oscillation in $M(H,n)$ was confirmed. 
Nakano\cite{nakano} and Alexandrov and Bratkovsky\cite{alex} account for 
the $\beta$-$\alpha$ oscillation 
in the magnetization $M(H,n)$ by using two independent energy band model 
neglecting the magnetic breakdown. 
Harrison et. al~\cite{harrison} 
calculated by 
the density of states 
derived from Pippard network model,~\cite{Pippard62} and 
found the $\beta$-$\alpha$ oscillation. 
In addition they\cite{nakano,alex,harrison} reveal that 
there is no $\beta$-$\alpha$ oscillation in the
magnetization $M(H,\mu)$ 
when the chemical potential, $\mu$ is fixed. 
In other words, they clarify that the quantum interference oscillation
such as the $\beta$-$\alpha$ oscillation is caused by the
chemical potential oscillation. 
In fact in our model, in the case of the fixed chemical 
potential, 
the $\beta$-$\alpha$ oscillation in $M(H,\mu)$ 
did not appear.\cite{kishigi3} 
Moreover, Sandu et al.\cite{sandu}and Han et al.\cite{so} found 
the $\beta$-$\alpha$ oscillation in $M(H,n)$ from the 
full-quantum-mechanical calculation 
by use of more realistic tight-binding model based on the 
band calculation. 

In our spinless tight-binding 
model,\cite{kishigi,kishigi2,kishigi3} 
the energy spectrum becomes 
so-called Hofstadter's butterfly diagram
\cite{Hofstadter} 
in the presence of magnetic field. 
Various oscillations with $f_\alpha$, $f_\beta$, $f_{2\beta}$, 
$f_{\beta -\alpha}$, etc.,
in $M(H,n)$ are due to the chemical potential oscillating.  
In that calculation the oscillation with $f_{\beta +\alpha}$ 
($\beta +\alpha$ oscillation) is much 
smaller than $\beta -\alpha$ oscillation, although the 
$\beta +\alpha$ oscillation is obtained to be larger than 
the $\beta -\alpha$ oscillation in the independent band 
model.\cite{nakano} The origin of the suppression of 
the $\beta +\alpha$ oscillation in the magnetic breakdown 
system is not clear yet, but since the $\beta +\alpha$ oscillation
is large in the case of the fixed chemical potential\cite{kishigi3}, 
a cancelation between the effects of the magnetic breakdown and 
the chemical potential oscillation will cause the suppression 
of the $\beta +\alpha$ oscillation. 
If we take account of the spin, 
the energy spectrum will be 
divided into 
two energy bands. 
Even if the total number of electrons is fixed, 
the electron number for each spin is not fixed. 
As a result we may expect a richer effect than the reduction 
factor in the semiclassical theory\cite{Shoenberg84} by 
taking account of the effect of spin in the system with 
the magnetic breakdown. 
However, it has never been studied that how the spin affects the 
``prohibited'' quantum interference 
oscillation ($\beta$-$\alpha$ oscillation) 
and quantum magnetic oscillations ($\alpha$, $\beta$, 
$\beta$+$\alpha$, $2\beta$ oscillations, etc.). 
Therefore, we need to investigate the effect of the spin 
on the magnetic breakdown and 
quantum interference oscillation. 

In the thermodynamics quantity such as the magnetization, 
$\beta$+$\alpha$ oscillation are 
observed in addition to 
$\alpha$, $2\alpha$, $\beta$-$\alpha$, $\beta$ and 2$\beta$ oscillations
in the quasi-two-dimensional organic 
conductors.\cite{Meyer,uji,honold} 
However, 
the peak at $f_{\beta -\alpha}$ reported by Meyer et al.\cite{Meyer} 
is 
very small, while 
that peak is large in the experiment by Uji et al.\cite{uji} 
Both de Haas van Alphen experiments measured at 
the same material, $\kappa$-(BEDT-TTF)$_2$Cu(NCS)$_2$, are done at 
the same field region ($22$ T $\sim$ $30$ T) and the 
same temperature ($\sim 0.4$ K). 
The difference in the amplitudes of $\beta -\alpha$ oscillation 
may be attributable to the effect of spin in a tilted magnetic field. 


In this paper, 
we calculate $M(H,n)$ from the energy, 
$E$, at the canonical ensemble 
by adding the Zeeman term to 
the previous spinless tight-binding 
model,\cite{kishigi,kishigi2,kishigi3} 
and try to analyze the effect of the spin on 
the magnetic breakdown and the quantum interference oscillation.

\section{Formulation}
Let us study the two-dimensional tight-binding model. 
The lattice spacings in $x$- and $y$-axes are $a$ and $b^{\prime}$, 
respectively. 
The Hamiltonian at $H=0$ is 
\begin{eqnarray}
\hat{\cal H}&=&\hat{\cal K}+\hat{\cal V}. \\
\hat{\cal K}&=&\sum_{i, j,\sigma}t_{i, j}C^{\dagger}_{i,\sigma} C_{j,\sigma}
=\sum_{{\bf k},\sigma}
C^{\dagger}_{{\bf k},\sigma} 
\epsilon({\bf k},\sigma)C_{{\bf k},\sigma}.\\ 
\hat{\cal V}&=&V\sum_{i,\sigma}\cos{\bf Q}\cdot {\bf
r}_iC^{\dagger}_{i,\sigma} C_{i,\sigma}. 
\end{eqnarray}
In the above $\sigma=\uparrow, \downarrow$, 
$C^{\dagger}_{i,\sigma}$ is the creation 
operator of $\sigma$ spin electron at $i$ site, 
$\epsilon({\bf k},\sigma)=-2t_a\cos ak_x-2t_b \cos b^{\prime}k_y$
and ${\bf Q}=(0,\pi /b^{\prime})$ 
is the lattice potential vector, where the Brillouin zone is 
$-\pi/a\leq k_x<\pi/a$, $-\pi/b^{\prime}\leq k_y<\pi/b^{\prime}$. 
In the momentum space, eq. (3) 
is written as 
\begin{eqnarray}
\hat{\cal V}&=&\frac{V}{2}\sum_{k_x,k_y^{\prime},m,\sigma}\{
C^{\dagger}(k_x,k_y^{\prime}+(m+1)\frac{\pi}{b^{\prime}},\sigma)
\nonumber  \\
&\times&C(k_x,k_y^{\prime}+\frac{m\pi}{b^{\prime}},\sigma) 
+h.c.\},
\end{eqnarray}
where summation in $k_y^{\prime}$ and $m$ are done in 
$-\pi/2b^{\prime}\leq k_y^{\prime}<\pi/2b^{\prime}$ 
and $m=0,1$, 
respectively. 
By $\hat{\cal V}$, the Brillouin zone is 
halved so that the first Brillouin zone becomes 
$-\pi/a\leq k_x<\pi/a$, $-\pi/b\leq k_y<\pi/b$, where 
we choose $b=2b^{\prime}$. 
In this paper, 
we take parameters as 2/3-filling, $t_b/t_a=1.0$ and $V/t_a=0.09$. 
Then, the 
Fermi surface is obtained 
as shown in Fig. 1, which has a characteristic topology 
in quasi-two-dimensional organic conductors.\cite{review,wos} 
From this Fermi surface we can see that 
the possible cyclotron orbits are 
the closed $\alpha$ orbit and 
the breakdown $\beta$ orbit.

When the magnetic field is applied parallel to $z$ axis, 
we take the Landau gauge $\bf{A}$=(0,$Hx$,0), and we 
write $\hat{\cal K}$ by Peierls substitution as 
\begin{eqnarray}
&\hat{\cal H}_{\rm p}&=\sum_{k_{x}^{\prime}, k_{y}^{\prime},n,m,\sigma}
\{-2t_a(\cos a(k_{x}^{\prime}+n\delta))-\frac{1}{2}g\mu_{\rm B}H\sigma\}
\nonumber  \\
&C&^{\dagger}(k_{x}^{\prime}+n\delta, k_{y}^{\prime}+m\frac{\pi}{b^{\prime}}, 
\sigma)
C(k_{x}^{\prime}+n\delta, k_{y}^{\prime}+m\frac{\pi}{b^{\prime}}, 
\sigma) \nonumber \\ 
&+&\{-t_b\exp[ib^{\prime}k_y^{\prime}+im\pi] 
C^{\dagger}(k_{x}^{\prime}+n\delta, k_{y}^{\prime}+m\frac{\pi}{b^{\prime}},
\sigma) \nonumber  \\
&C&(k_{x}^{\prime}+(n+1)\delta, k_{y}^{\prime}+m\frac{\pi}{b^{\prime}}, 
\sigma)+h.c.\}, 
\end{eqnarray}
where $\delta=eaH/\hbar c=(\phi/\phi_{0})(2\pi /a), \phi=ab^{\prime}H$ 
is the flux passing
through a unit cell 
and $\phi_0=2\pi\hbar c/e$ is a unit flux, and $h=\phi/\phi_0$ is the 
number of the flux quantum per unit cell. We 
represent magnetic fields by $h$, henceforth. 
When $h$ is a rational number, namely, $h=\phi/\phi_{0}=p/q$ with
$p$ and $q$ 
being mutually prime integers, 
the matrix size of $\hat{\cal H}_{\rm p}$ is $q\times q$. 
In eq. (5), 
$0 \leq k_{x}^{\prime}
<\delta /p$ and 
$0\leq n<q$. 
As a result, the matrix size of 
$\hat{\cal H}_{\rm p}$+$\hat{\cal V}$ becomes $2q\times 2q$. 
In $\hat{\cal H}_{\rm p}$, 
$\mu_{\rm B}$ is the Bohr magneton and $g\simeq 2$. 
If the Zeeman term in $\hat{\cal H}_{\rm
p}$+$\hat{\cal V}$ is excluded, this model becomes the previous spinless 
model.\cite{kishigi,kishigi2,kishigi3}

Under the condition of the fixed total electron number, $n$, 
the ground state energy per site is calculated by 
\begin{eqnarray}
E(h,n)=\frac{1}{N_s}\sum^{n}_{j=1}\epsilon_{j}, 
\end{eqnarray}
where $N_s$ is the total site number and 
$\epsilon_{j}$ is one electron 
eigenvalue, and index $j$ includes the spin index. 
The magnetization is given by 
$M(h,n)=-\partial E(h,n)/\partial h$.

The magnetic field cannot be changed continuously in this
formulation
and the experimentally accessible field ($H\sim 10$ T) 
is difficult to study due to the
large value of $q$. For example, as $a=12.7\AA$ and $b=8.4\AA$ for
(BEDT-TTF)$_2X$,\cite{review,wos}
$h=p/q$ is about 1/400 when $H\simeq 10$ T. 
We calculate at 
higher field ($H\sim 100$ T, $h\sim 1/40$).

In order to see the effect of the Zeeman term clearly, 
we define $\widetilde{g}$ as $\widetilde{g}h\equiv
\frac{1}{2}g\mu_{\rm B}H/t_a$. 
For example, in quasi-two-dimensional organic conductors with 
the transfer integral, $t_a=250 k_{\rm B}$,\cite{review,wos} and 
the lattice spacings mentioned above ($a=12.7\AA$ and $b=8.4\AA$) 
we get 
$\widetilde{g}\simeq g \simeq 2.0$. 

\begin{figure}
\leavevmode
\epsfxsize=8cm
\epsfbox{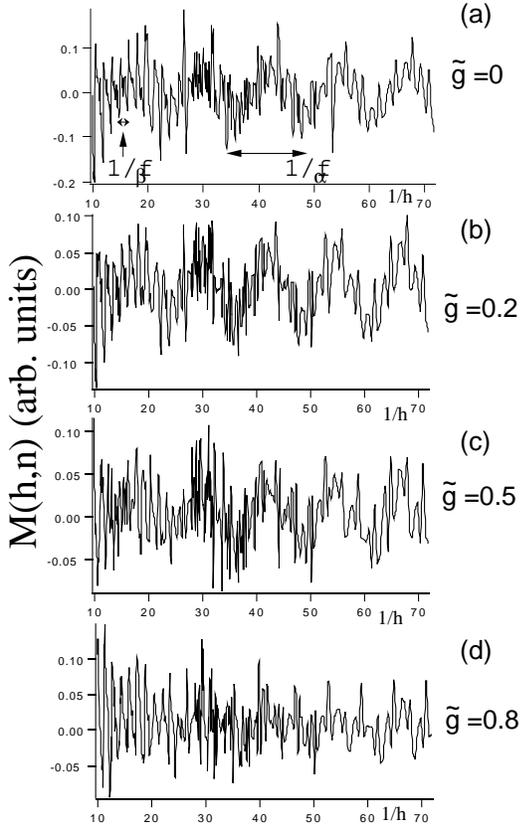}
\caption{
Magnetization as a function of the inverse field. 
}
\end{figure}

\section{Result and Discussion}

We show $M(h,n)$ 
at $\widetilde{g}=0, 0.2, 0.5$ and 0.8 
in Fig. 2. 
and the Fourier transform of $M(h,n)$ 
in Fig. 3. 

First, we see Figs. 2(a) and 3(a) when $\widetilde{g}=0$. 
In Fig.3(a), these peaks 
($f_\alpha \simeq 0.08$, $f_{2\alpha} \simeq 0.16$, 
$f_{4\alpha} \simeq 0.32$, 
$f_\beta \simeq 0.67$, $f_{2\beta} \simeq 1.33$ and 
$f_{\beta -\alpha} \simeq 0.59$)
correspond to the area of 
$\alpha$, $2\alpha$, $4\alpha$, $\beta$, $2\beta$ and
$\beta$-$\alpha$ orbits\cite{orbits} 
(the ratios of area of $\alpha$ and $\beta$ 
orbits to the first Brillouin zone are 
0.08 and 0.67, respectively). 
The oscillations of the long 
period ($f_\alpha$) and that of the short period ($f_\beta$) 
in $M(h,n)$ are seen in Fig. 2(a). 
As $h$ increases, the amplitude of the oscillation with 
$f_\beta$ becomes gradually larger. 
This behavior is due to the magnetic breakdown.
Although 
the $\beta$-$\alpha$ oscillation in addition to $\alpha$, $\beta$ and 
$2\beta$ oscillations exists, a peak for 
$\beta$+$\alpha$ oscillation is very small. 
This property has been shown previously by one of the 
authors,
\cite{kishigi,kishigi2,kishigi3} 
and Harrison et al\cite{harrison} also have 
obtained the similar result.

\begin{figure}
\leavevmode
\epsfxsize=8cm
\epsfbox{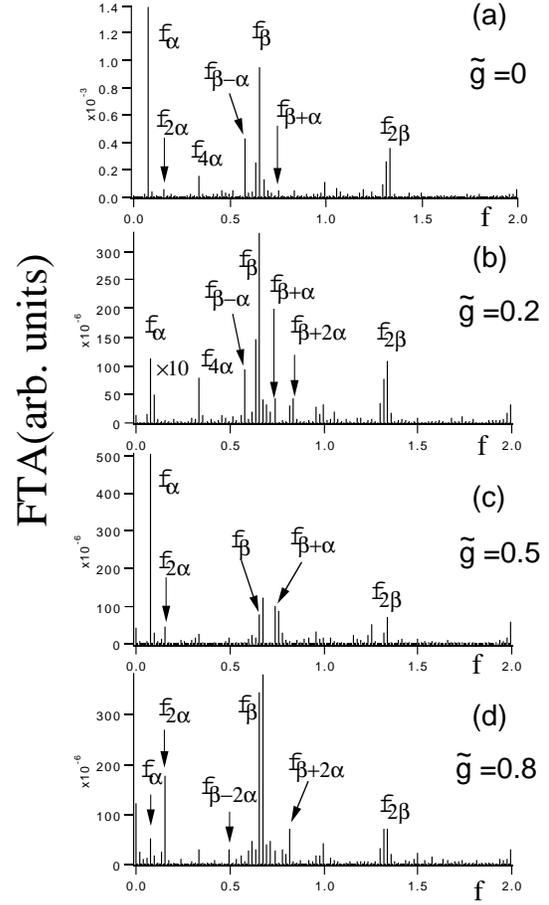}
\caption{The Fourier transform amplitude (FTA) of $M(h,n)$. 
The range of the Fourier transform is 
$10\leq h^{-1}\leq 60$.
}
\end{figure}

Next, from the Fourier transform of $M(h,n)$ for $\widetilde{g}\neq0$ 
(Fig. 3(b)$\sim$(d)), 
we can see that 
the peak at $f_{\beta +\alpha}$ 
becomes large at $\widetilde{g}=0.2$, 
and 
this peak is the same order of magnitude as 
at 
$f_\beta$, whereas the peak at $f_{\beta -\alpha}$ 
is very small at $\widetilde{g}=0.5$. 
When $\widetilde{g}=0.8$, 
peaks at $f_{\alpha}$ and $f_{\beta \pm\alpha}$ 
become very small, 
whereas peaks at $f_{2\alpha}$ and $f_{\beta \pm2\alpha}$ become 
large, where the $\beta -2\alpha$ oscillation is the 
quantum interference oscillation like $\beta -\alpha$ oscillation. 
In order to understand the $\widetilde{g}$-dependence of these 
peaks, we show amplitudes of these peaks at 
$f_\alpha$, $f_{2\alpha}$, $f_\beta$, 
$f_{\beta \pm\alpha}$ and
$f_{\beta \pm2\alpha}$ as a function of $\widetilde{g}$ 
from $\widetilde{g}=0$ to 1.0 in Fig. 4. 


\begin{figure}
\leavevmode
\epsfxsize=8.7cm
\epsfbox{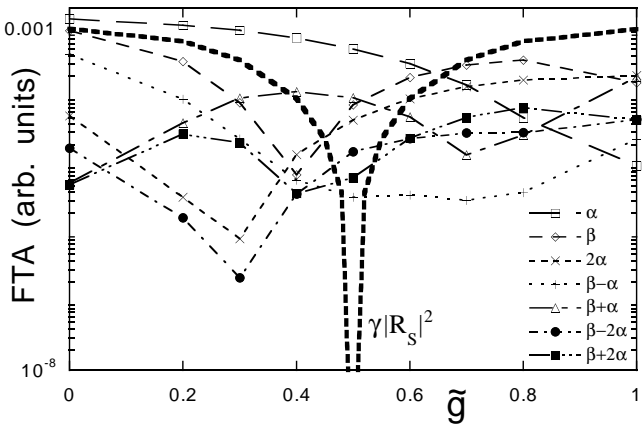}
\caption{The FTAs of 
some peaks ($f_\alpha$, $f_{2\alpha}$, $f_\beta$, 
$f_{\beta \pm\alpha}$ and 
$f_{\beta \pm2\alpha}$). 
Thick Dotted lines indicate $\gamma|R_{\rm s}|^2$. 

}
\end{figure}


In
semiclassical theory, 
the effect of the spin is represented 
by the 
reduction factor\cite{Shoenberg84}, 
$R_{\rm s}=\cos (\frac{1}{2}p\pi g m/m_{0})$, 
where $p$ is for $p$th harmonics of the each frequency in 
$M(h,n)$ and $m$ ($m_0$) are cyclotron effective mass (free electron mass). 
For example, when we choose $m/m_0=2$, $p=1$, 
$t_a=250 k_{\rm B}$, $a=12.7\AA$ and $b=8.4\AA$,  
the Fourier transform amplitude (FTA) of $M(h,n)$ is 
reduced by the factor 
$|R_{\rm s}|^2=|\cos (\pi \widetilde{g})|^2$. 
We show $\gamma|R_{\rm s}|^2$ in Fig. 4, where 
we take $\gamma=0.001$. 
From Fig. 4, the $\widetilde{g}$-dependences of the 
FTAs of 
$f_\alpha$, $f_{2\alpha}$, $f_\beta$, 
$f_{\beta -\alpha}$ and $f_{\beta -2\alpha}$ 
decrease as $\widetilde{g}$ increases, and 
all of these except 
$f_\alpha$ increase upon 
further increasing 
$\widetilde{g}$. 
These FTAs are in agreement with 
$|R_{\rm s}|^2$ qualitatively. 
It may be due to the different values of 
the effective masses and $p$ 
that the position of the minimum of the FTAs of these frequencies 
are different. 
However, note that the $\widetilde{g}$-dependences of 
the FTAs of $f_{\beta +\alpha}$ 
and $f_{\beta +2\alpha}$ 
are quite different from $|R_{\rm s}|^2$; 
these amplitudes first increase as $\widetilde{g}$ increases and 
decrease as $\widetilde{g}$ increases further. 

The anomalous $\widetilde{g}$-dependences can be understood 
as follows. In the spinless case ($\widetilde{g}=0$), 
amplitudes of $\beta +\alpha$ 
and $\beta +2\alpha$ oscillations are suppressed by the cancelation of the 
magnetic breakdown and the chemical potential oscillation. 
By finite $\widetilde{g}$ the chemical potential oscillation is strongly
suppressed. The suppression of the chemical potential oscillation 
is consistent with a strong reduction of 
$\beta -\alpha$ and $\beta -2\alpha$ oscillations in 
small $\widetilde{g}$ as seen in Fig. 4. 
On the other hand, the magnetic breakdown is not affected 
by $\widetilde{g}$. As a result, $\beta +\alpha$ and 
$\beta +2\alpha$ 
oscillations becomes large as 
$\widetilde{g}$ increases due to an incomplete 
cancelation of the chemical potential oscillation 
and the magnetic breakdown. 

When the field is tilted from 
the $k_z$ axis to
the $k_x$-$k_y$ plane by $\theta$, 
the component of 
the magnetic field for the Fermi surface along the $k_z$ axis 
is $H\cos \theta$. 
That for the Zeeman term is not changed by tilting of $H$. 
Therefore, 
the effect of magnetic fields 
for the Zeeman term 
is enhanced 
by tilting the field. 


As seen in Fig. 4 the amplitude of 
$\beta -\alpha$ oscillation 
depends on $\widetilde{g}$, i.e., the tilting angle $\theta$. 
Therefore, we expect that the different results 
between Meyer et al.\cite{Meyer} and Uji et al.\cite{uji} are 
due to a direction of magnetic fields 
in both experiments. 
In fact, the frequency of $\beta$ oscillation 
in both experiments are not the same.\cite{Meyer,uji} 
The relative magnitudes of the 
amplitude of each frequencies ($f_{\alpha}$, $f_{\beta}$, 
$f_{\beta -\alpha}$, $f_{\beta +\alpha}$, etc.) 
are varied 
by changing $\widetilde{g}$. 
Particularly, the oscillations with $f_{\beta +\alpha}$ 
and $f_{\beta +2\alpha}$ have the anomalous 
$\widetilde{g}$-dependences. 
Thus, we should consider the 
system of the magnetic breakdown with the effect of the spin, which 
has not been taken into account in the previous 
studies.\cite{kishigi,kishigi2,kishigi3,nakano,alex,harrison,sandu,so}


\section{Conclusion}

From the full-quantum mechanical calculation by using 
the simple tight-binding model including an electron spin, 
we analyze the effect of the spin on the quantum interference 
oscillation such as $\beta -\alpha$ and $\beta -2\alpha$ 
oscillations and 
the quantum magnetic oscillation such as 
$\alpha$, $2\alpha$, $\beta$, $\beta +\alpha$ 
and $\beta +2\alpha$ oscillations. 
It is obtained that 
the relative magnitudes of all oscillations 
($\alpha$, $\beta$, $\beta +\alpha$, $\beta -\alpha$, 
etc.) are changed by the Zeeman term. 
In particular, the amplitudes of 
$\beta +\alpha$ and $\beta +2\alpha$ oscillations 
have 
anomalous $\widetilde{g}$-dependences. 

We expect that 
the $\widetilde{g}$-dependences of the amplitudes of these oscillations 
in our result will be observed in 
the experiment of tilting magnetic field.



\section{Acknowledgment}

The authors thank 
M. Nakano for valuable discussions. 
One of the authors (K. K.) thanks S. Uji for useful discussions.  
This work was partially supported by Grant-in-Aid for
JSPS Fellows from the Ministry of
Education, Science, Sports and Culture.
K. K. 
was financially supported by the Research Fellowships
of the Japan Society
for the Promotion of Science for Young Scientists.

\end{document}